# On the origin of field-like spin-orbit torques in heavy metal/ferromagnet/oxide thin film heterostructures

Yongxi Ou[1], Chi-Feng Pai[*,1], Shengjie Shi[1], D. C. Ralph[1,2], and R. A. Buhrman[1,+]

[1]Cornell University, Ithaca, New York 14853, USA

[2]Kavli Institute at Cornell, Ithaca, New York 14853, USA

We report measurements of the thickness and temperature ($T$) dependencies of current-induced spin-orbit torques, especially the field-like (FL) component, in various heavy metal (HM)/normal metal (NM) spacer/ferromagnet (FM)/Oxide (MgO and HfO$_x$/MgO) heterostructures. The FL torque in these samples originates from spin current generated by the spin Hall effect (SHE) in the HM. For a FM layer sufficiently thin that a substantial portion of this spin current can reach the FM/Oxide interface, $T$-dependent spin scattering there can yield a strong FL torque that is, in some cases, opposite in sign to that exerted at the NM/FM interface.

[*] Present address: Department of Materials Science and Engineering, National Taiwan University, Taipei 10617, Taiwan

Since the discovery that an in-plane charge current density $J_e$ in certain HM thin films can be utilized to effectively manipulate the magnetization state of an adjacent FM layer [1–6], two classes of spin orbit interactions mechanisms have been considered as candidate sources for these spin-orbit torques (SOT): a strong Rashba-Edelstein (RE) interaction [1,2,7–9] at the ferromagnet interface(s) resulting in the generation of a non-equilibrium conduction election spin polarization at the interface; and the effect of an incident transverse spin current density $J_s$ on the ferromagnet arising from a strong spin Hall effect (SHE) within certain HMs [3–6,10]. Both mechanisms can result in the exertion of damping-like (DL) torque $\tau_{DL}$ and a FL torque $\tau_{FL}$ on the FM. These can be characterized by the DL(FL) spin torque efficiency $\xi_{DL}(\xi_{FL})$ or the equivalent effective fields generated per unit $J_e$, $\Delta H_{DL}/\Delta J_e = \xi_{DL}\hbar/(8\pi e M_s t_{FM})$ and $\Delta H_{FL}/\Delta J_e = \xi_{FL}\hbar/(8\pi e M_s t_{FM})$, where $M_s$ is the saturation magnetization of the FM layer, $t_{FM}$ is its thickness, $\hbar$ is Planck's constant and $e$ is the electron charge [10,11]. The RE effect is generally expected, at least within the context of a Boltzmann equation or drift-diffusion analysis [10], to exert a larger FL than DL torque, while in the SHE case absorption of the transverse polarized component of $J_s$ exerts a larger $\tau_{DL}$ and reflection with some spin rotation can result in a smaller $\tau_{FL}$.

Studies of SOT excitation of nanomagnets and domain wall motion in HM/FM heterostructures have generally shown that these processes can be well explained by a DL torque due to the SHE of the HM, with an interfacial Dzyaloshinskii-Moriya interaction also important in the case of domain wall displacement [12–14]. It is therefore quite puzzling that heterostructures made of the same materials can, when the FM is thin and magnetized out of plane, exhibit $\tau_{FL} > \tau_{DL}$ [12,15]. This is true even though experiments in which a NM layer of variable thickness (with minimal SHE) is inserted between the HM and FM, confirm the nonlocal nature of $\tau_{FL}$ in those experiments [16–18]. The origins of $\tau_{FL}$ are therefore under active debate -- there are reports showing that the magnitude and even the sign of $\tau_{FL}$ can greatly depend on the thickness of the FM [17,19], the type of FM [20], the type of HM [16,21], the direction of the magnetization in FM [22,23] and temperature [15,23].

Here we report measurements of SOTs in various in-plane magnetized (IPM) and perpendicularly magnetized (PM) HM/NM/FM/Oxide (MgO and HfO$_x$/MgO) heterostructures, using Ta, Pt, and W for the HM. We have examined $\tau_{FL}$ as a function of NM and FM thickness, $t_{NM}$ and $t_{FM}$, and as a function of temperature $T$ from 300 K to 5 K. The spin torque efficiencies, $\xi_{DL}$ and $\xi_{FL}$, are measured by spin torque ferromagnetic resonance (ST-FMR) for IPM samples [3,24] and $\Delta H_{DL}$ and $\Delta H_{FL}$ by the harmonic response (HR) method [19,25,26] for PM samples. The $t_{NM}$ dependent measurements reveal that the FL torques observed in all of our samples are due to spin current that originates in the HM via the SHE. By varying $t_{FM}$ for PM cases we find that the FL torque in samples with very thin FM layers can be strong, and differ significantly between FM/MgO and FM/HfO$_x$/MgO interfaces. Whereas the $\tau_{DL}$ is invariably only weakly sensitive to $T$, and similarly for the relatively smaller $\tau_{FL}$ present in samples with a thick FM layer, the FL contribution from the FeCoB/Oxide interface is strongly $T$ dependent, seemingly going to zero at low $T$ in a quasi-linear manner, consistent with previous measurements [15,23].

We will first establish that $\tau_{FL}$ in our HM/NM/FM/Oxide heterostructures is caused by $J_s$ that originates in the HM, rather than by a local RE effect at a FM interface. The IPM multilayer samples used in this study (see supplementary material [27] for sample fabrication and processing details) were W(4)/Hf($t_{Hf}$)/Fe$_{60}$Co$_{20}$B$_{20}$($t_{FeCoB}$)/MgO(2)/Ta(1), where $t_{Hf} = 0.25 - 2$ nm and $t_{FeCoB} = 2 - 7$ nm, and the numbers in parentheses represent the nominal thickness in nm. We used ST-FMR to define a FMR spin torque efficiency $\xi_{FMR}$ from the ratio of the symmetric (*S*) and anti-symmetric (*A*) components of anisotropic magnetoresistance response at the ferromagnetic resonance [24,28]. *S* is proportional to $\tau_{DL}$ and *A* is due to the sum of the Oersted field torque and any spin-orbit-induced $\tau_{FL}$. The spin-orbit torque efficiencies $\xi_{DL}$ and $\xi_{FL}$ can then be obtained from the dependence of $\xi_{FMR}$ on $t_{FM}$, assuming that $\xi_{DL}$ and $\xi_{FL}$ do not have a strong dependence on $t_{FM}$ in the range examined [24,28]:

$$\frac{1}{\xi_{FMR}} = \frac{1}{\xi_{DL}}\left(1 + \frac{\hbar}{e}\frac{\xi_{FL}}{4\pi M_s t_{FM} t_{HM}}\right) \qquad (1)$$

Here $t_{HM}$ is the thickness of the HM. Figure 1(a) shows as an example the results for $\xi_{FMR}$ as a function of FeCoB thickness for two different Hf spacer thicknesses. The strong variation with $t_{FM}$ is indicative of a significant $\tau_{FL}$. From fits to Eq. (1) (dashed lines in Fig. 1(a)) we determined $\xi_{DL}$ and $\xi_{FL}$ as a function of $t_{Hf}$, with the results shown in Fig. 1(b). The signs of both $\xi_{DL}$ and $\xi_{FL}$ are negative, indicating a negative spin Hall ratio for W and a FL effective field that is in opposition to the current-generated Oersted field. As $t_{Hf}$ increases from 0.25 nm, we find that both $|\xi_{DL}|$ and $|\xi_{FL}|$ decrease in concert, and extrapolate to negligible values at large Hf thicknesses[2]. The solid and dashed lines in Fig 1(b) are fits of the same spin current attenuation function to both sets of data [27], which indicate an effective attenuation length $\lambda_s^{Hf} \approx 0.9$ nm. This indicates that both $\xi_{DL}$ and $\xi_{FL}$ are the result of a diffusive spin current from the underlying W that passes through the Hf layer. Similar non-local FL terms have been reported for Py/Cu/Pt and CoFeB/Cu/Pt IPM heterostructures [16,17]. The fit for $\xi_{FL}$ suggests that there is a small residual FL component ($-0.010 \pm 0.004$) when the Hf is in the thick limit, which may be attributable to a comparably weak RE effect at the Hf/FeCoB interface [29]. In contrast to the results that we will discuss for PM samples with much thinner FM layers, for the IPM samples neither $\xi_{DL}$ nor $\xi_{FL}$ has a significant $T$ dependence [27].

---

[2] While there are reports that Hf can sometimes generate a quite substantial spin current [29,35], our amorphous Hf layers have a minimal SHE, consistent with a previous experiment [31]. We tentatively attribute this difference to the different phases that Hf thin films can possess, as in the case of W.

The FL torque can be considerably stronger in PM samples, which have by necessity much thinner FM layers. We will show that the degree of this enhancement is strongly dependent on both $t_{FM}$ and the details of the FM/Oxide interface. To establish the first dependency we fabricated two series of PM samples: Ta(4)/FeCoB($t_{FeCoB}$)/MgO and Ta(6)/FeCoB($t_{FeCoB}$)/Hf($t_{Hf}$) (oxidized)/MgO. In the latter case, $0 < t_{Hf} \leq 0.4$ nm, and the Hf was oxidized to form a thin HfO$_x$ layer during the subsequent sputter deposition of the MgO rather than the usual case of oxidation of the FM surface [30], as confirmed by the interfacial anisotropy energy, $K_s \approx 2$ erg/cm$^2$, that was obtained without any post-fabrication annealing [27]. Figure 2(a) shows the effective fields as measured by HR as a function of $t_{FeCoB}$, for $t_{Hf} = 0$ and 0.2 nm. (Note that all HR results reported in this paper have been corrected for the transverse (planar Hall) magnetoresistance contribution to the HR measurement [26,27], and also that the small Oersted field contribution has been subtracted from the $\Delta H_{FL} / \Delta J_e$ results.) For both the FM/MgO and FM/HfO$_x$/MgO cases we find that $\Delta H_{FL} / \Delta J_e$ decreases rapidly as $t_{FM}$ increases from 0.7 nm to 1.0 nm. For a given value of $t_{FM}$, the strength of $\tau_{FL}$ is very different -- stronger for the FM/HfO$_x$/MgO samples by approximately a factor of 2 compared to FM/MgO.

Even though $\Delta H_{FL}/\Delta J_e$ depends strongly on both $t_{FM}$ and the composition of the FM/Oxide interface for the PM samples, our previous studies as a function of the thickness of a Hf spacer between the HM and the FM in PM samples show that the origin of this enhanced $\tau_{FL}$ is still a spin current emitted from the HM, and the spin-orbit torque decay as a function of increasing Hf spacer thickness and extrapolate to negligible values at large $t_{Hf}$ [31]. Therefore we conclude that the enhanced FL torque in the PM samples must be due to the portion of $J_s$ from the HM that can pass through the FM layer and reach the FM/Oxide interface before dephasing and/or relaxing, so that spin scattering at this interface is able to affect the amount of spin accumulation in the very thin FM layer.

As noted above, for a given $t_{FM}$, $\Delta H_{FL}/\Delta J_e$ for FeCoB/HfO$_x$ is approximately twice that for FeCoB/MgO. The importance of the FM/Oxide interface in the enhancement of $\tau_{FL}$ in PM structures is also illustrated by measurements of $\Delta H_{DL}/\Delta J_e$ and $\Delta H_{FL}/\Delta J_e$ as a function of the thickness of an oxidized Hf passivation layer for a series of Ta(6)/FeCoB(0.8)/Hf($t_{Hf}$)/MgO. As $t_{Hf}$ increases from approximately one atomic layer (0.2 nm before oxidation) to two (0.4 nm) there is only a small change in $\Delta H_{DL}$, while $\Delta H_{FL}$ decreases markedly, by nearly a factor of two (Fig. 2(b)). The strong variation of the FL term with $t_{Hf}$ is presumably related to the more complete passivation of the FM surface by a slightly thicker Hf layer. Our finding that the perpendicular anisotropy field $H_a$ increases as $t_{Hf}$ becomes thicker, as shown in the inset to Fig.

2(c), supports this attribution. In Fig. 2(c) we plot $\Delta H_{FL}/\Delta J_e$ as a function of $1/H_a$, which indicates that as the strength of the interfacial anisotropy increases, $\tau_{FL}$ decreases linearly. This behavior is repeated with Ta/FeCoB/MgO samples where we found that $\Delta H_{FL}/\Delta J_e$ for that system also varies as $1/H_a$ when different annealing temperatures were employed to modify the PMA [27]. This suggests a tradeoff between the SOI that creates the PMA at FCB/Oxide interfaces [32] and the spin relaxation mechanism responsible for $\tau_{FL}$.

While there is only a weak $T$ dependence for $\xi_{DL}$ and $\xi_{FL}$ in the IPM samples, for the thin PM samples, as we will show, the FL torque contribution from the FM/Oxide interface is strongly temperature dependent, weakening dramatically at low $T$. We tentatively ascribe this behavior to $T$-dependent spin-flip scattering at the FM/Oxide interface. When the spin scattering is suppressed near $T = 0$, the enhancement of FL torque generated by the FM/Oxide interface is suppressed as well, even though the DL torque is largely unaffected.

We performed $T$-dependent HR measurements of $\Delta H_{DL}/\Delta J_e$ and $\Delta H_{FL}/\Delta J_e$ on a number of different PM samples. Fig. 3(a) and 3(b) show results obtained respectively from an annealed Ta(4)/FeCoB(0.8)/MgO sample and an un-annealed Ta(6)/FeCoB(0.8)/Hf(0.2) (oxidized)/MgO sample, that is without and with the Hf passivation layer. While $\Delta H_{DL}/\Delta J_e$ is nearly invariant with $T$, there is a strong variation in $\Delta H_{FL}/\Delta J_e$ in both cases as $T$ goes towards zero. The behavior of $\Delta H_{FL}/\Delta J_e$ for the sample without the Hf passivation is quite similar to that reported previously [15,23]. Below 250 K, $\Delta H_{FL}$ decreases quasi-linearly with decreasing $T$, approaching zero around 70 K, and then departs from linearity to vary more slowly, becoming slightly negative as $T$ goes to zero. Here we use the convention that a negative $\Delta H_{DL}$ corresponds to a negative spin Hall angle. While $\Delta H_{FL}$ at 300 K decreased by over a factor of five when we increased $t_{FeCoB}$ from 0.7 to 1.0 nm in the Ta samples without Hf (Fig. 2 (a)), the results all converge to a very similar small, negative, value as $T$ approaches 5 K (see [27] plot S6).

The FL torque in the Hf passivated sample (Fig. 3(b)) also shows a strong $T$ dependence that is even more linear for $T$ below 200 K to at least 50 K. For additional insight we also studied PM samples with different HM base layers and with a thin Hf spacer layer between the HM and the FM: Ta,Pt,W(4)/Hf($t_{Hf}$)/Fe$_{60}$Co$_{20}$B$_{20}$($t_{FeCoB}$)/MgO(2)/Ta(1), where $t_{Hf} = 0.5$ or 1 nm and $t_{FeCoB} = 0.7 - 1$ nm. As mentioned previously we initially employed such Hf spacer layers to enhance the PMA in the HM/Hf/FeCoB/MgO heterostructures and to study the role of the spin current from the HM in determining the strength of the SOT [31]. Subsequently, analytical electron microscopy studies [33] have shown that there is diffusion of Hf to the bottom of the MgO overlayer, both before and more so after annealing, resulting in the formation of a thin HfO$_x$ layer at the top of the FM. We conclude that this FM/HfO$_x$ interface is responsible for the linear $T$ dependence of the FL torque in all these samples at low $T$.

Figure 3(c) shows $\Delta H_{DL}(T)$ and $\Delta H_{FL}(T)$ as generated in a Ta and a Pt based sample with the Hf spacer and Fig. 3(d) shows results from a W sample ($t_{Hf} = 1$ nm for the Ta and W samples and $t_{Hf} = 0.5$ nm for the Pt sample). $\Delta H_{DL}$ in all three cases is nearly constant, increasing just slightly with decreasing $T$. The different signs for $\Delta H_{DL}$ correlate with the different signs of the spin Hall ratio $\theta_{SH}$ (negative for Ta and W, positive for Pt). The sign of $\Delta H_{FL}(T)$ at $T = 300$ K is in all cases opposite to that of $\theta_{SH}$, being + for Ta and W, and – for Pt.

In all cases $\Delta H_{FL}(T)$ also decreases quite linearly with decreasing $T$ down to ~ 5 K and in the process exhibits a sign change at low $T$ for Ta and W, one that is most strongly seen in the W sample, which has the highest spin Hall ratio. This provides the strongest incident spin current at the Hf/FM interface and results in the largest field-like torque being exerted there, in comparison to the Ta/Hf/FM and Pt/Hf/FM samples. $\Delta H_{FL}(T)$ at the lowest temperature for the W sample corresponds to a FL spin torque efficiency $\xi_{FL} \approx -0.03$, which is consistent with the IPM ST-FMR measurement for the same structure with the same Hf thickness (Fig.1 (b)).

To further confirm that the $\Delta H_{FL}(T)$ behavior in these latter samples cannot be the result of a strong RE generated effective field at the Hf/FeCoB (or FeCoB/HfO$_x$) interface, we measured $\Delta H_{DL}(T)$ and $\Delta H_{FL}(T)$ in a control sample having only a 4 nm Hf base layer, as also shown in Fig. 3(d). $\Delta H_{DL}(T)/\Delta J_e$ is negligible over the full $T$ range, while $|\Delta H_{FL}(T)/\Delta J_e|$ is quite small, $\leq 1 \times 10^{-6}$ Oe/(A/cm$^2$) with little $T$ variation.

We conclude that the dominant mechanism in generating $\Delta H_{FL}$ is the scattering of the incident SHE-generated $J_s$ at each of the two interfaces of the FM. For the spin torque that is exerted only at the NM/FM interface (*e.g.*, in the IPM samples) the result is a comparatively small field-like torque, $\Delta H_{FL} < \Delta H_{DL}$, and is at most weakly $T$ dependent. When in thin PM samples a significant $J_s$ reaches the FeCoB/Oxide interface the result is a stronger contribution to $\tau_{FL}$ at $T \sim 300$ K with a sign opposite to $\theta_{SH}$ and with $\Delta H_{FL} \geq \Delta H_{DL}$, but this contribution decreases apparently close to zero at low $T$, leaving only the weaker spin current dependent contribution to $\Delta H_{FL}$ from the HM/FM interface.

The spin scattering at the HM/FM interface can perhaps be treated via the scattering-matrix spin mixing conductance scenario where spin rotation during the reflection of part of $J_s$ will result in a field-like torque. Initial calculations indicated that this effect should be weak, but until recently such calculations assumed no strong SOI at the interface, which is generally not the case in the HM/FM thin film systems of interest here [34]. Typically we find experimentally that $\Delta H_{FL} \sim 0.2 - 0.3\ \Delta H_{DL}$ [24]. At the FM/Oxide interface, the strength of $H_a$ is inversely correlated with the strength of the field-like torque (see Fig. 2(c) and [27], Fig. S4), which indicates that there may be competition between the electronic states at the FM/Oxide interface that generate the interfacial anisotropy and those that provide the spin relaxation pathway.

In summary, we have measured the thickness and *T* dependence of the DL and FL SOTs in a range of HM/NM/FM/Oxide heterostructures. The strength of the FL torque varies in a manner principally dependent upon: (1) a nonlocal $J_s$ generated by the SHE in the HM, and (2) scattering of this $J_s$ at the FM interfaces on which it impinges, rather than dependent on a large local RE effect in which a spin polarization is generated by the $J_c$ flowing at a FM interface. We infer that spin scattering at either the HM/FM or FM/Oxide interface can promote formation of a net spin accumulation in the FM layer that generates an effective field $\Delta H_{FL}$. The FM/Oxide interface can result in the stronger $\Delta H_{FL}$, provided that a significant portion of the spin current incident on the FM reaches that interface prior to dephasing, which requires a very thin FM layer, and provided that there is a high density of interfacial states that act as strong spin scattering centers. This latter appears to vary inversely with the interfacial anisotropy energy density in the FeCoB/Oxide interfaces studied here.

**Acknowledgement**

We thank Graham Rowlands and Junbo Park for helpful discussions and assistance on low $T$ measurements. This research was supported in part by ONR and by NSF/MRSEC (DMR-1120296) through the Cornell Center for Materials Research (CCMR), and NSF through use of the Cornell Nanofabrication Facility (CNF)/NINN (ECCS-1542081) and the CCMR facilities.

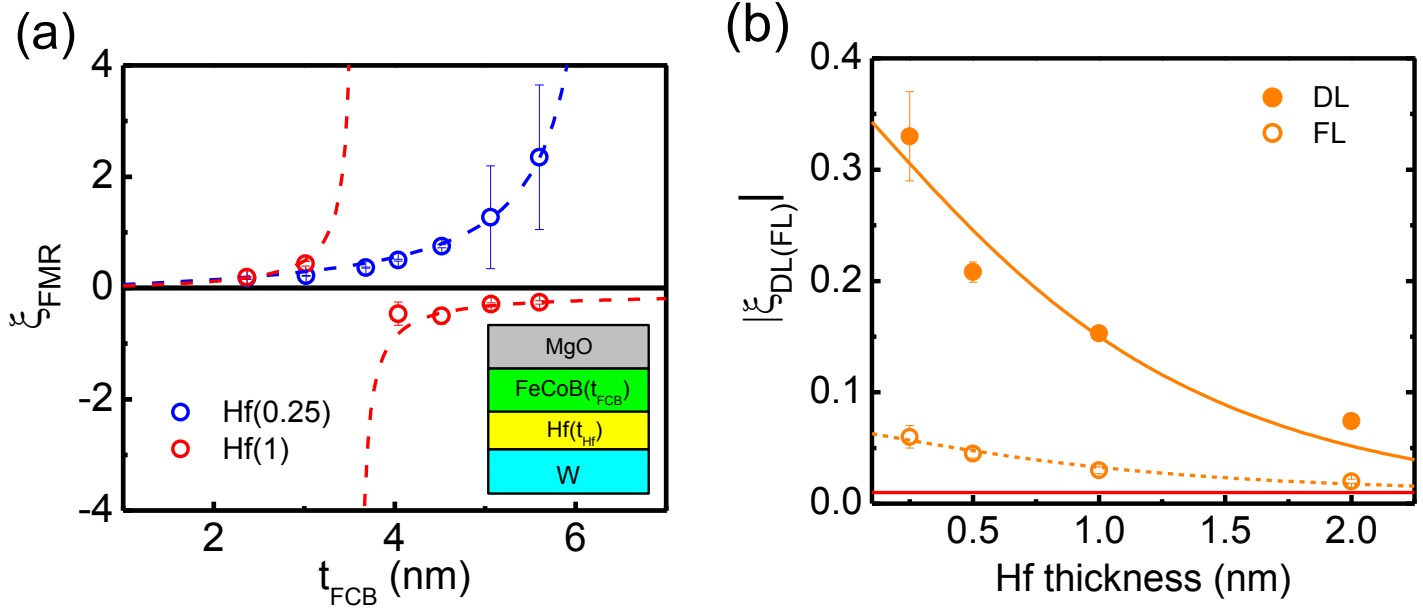

Fig. 1.(a) ST-FMR spin torque efficiency $\xi_{FMR}$ as a function of the FeCoB thicknesses $t_{FCB}$ for two in-plane magnetized samples W(4)/Hf(0.25,1)/FeCoB/MgO. (b) Damping-like and field-like spin torque efficiencies $\xi_{DL}$ and $\xi_{FL}$ as a function of Hf thickness. The solid (dashed) orange lines are fits to the two sets of data (see [27]). The sign of both $\xi_{DL}$ and $\xi_{FL}$ are negative. The red solid line indicates a small residual FL spin torque efficiency ($-0.010 \pm 0.004$).

Fig 2

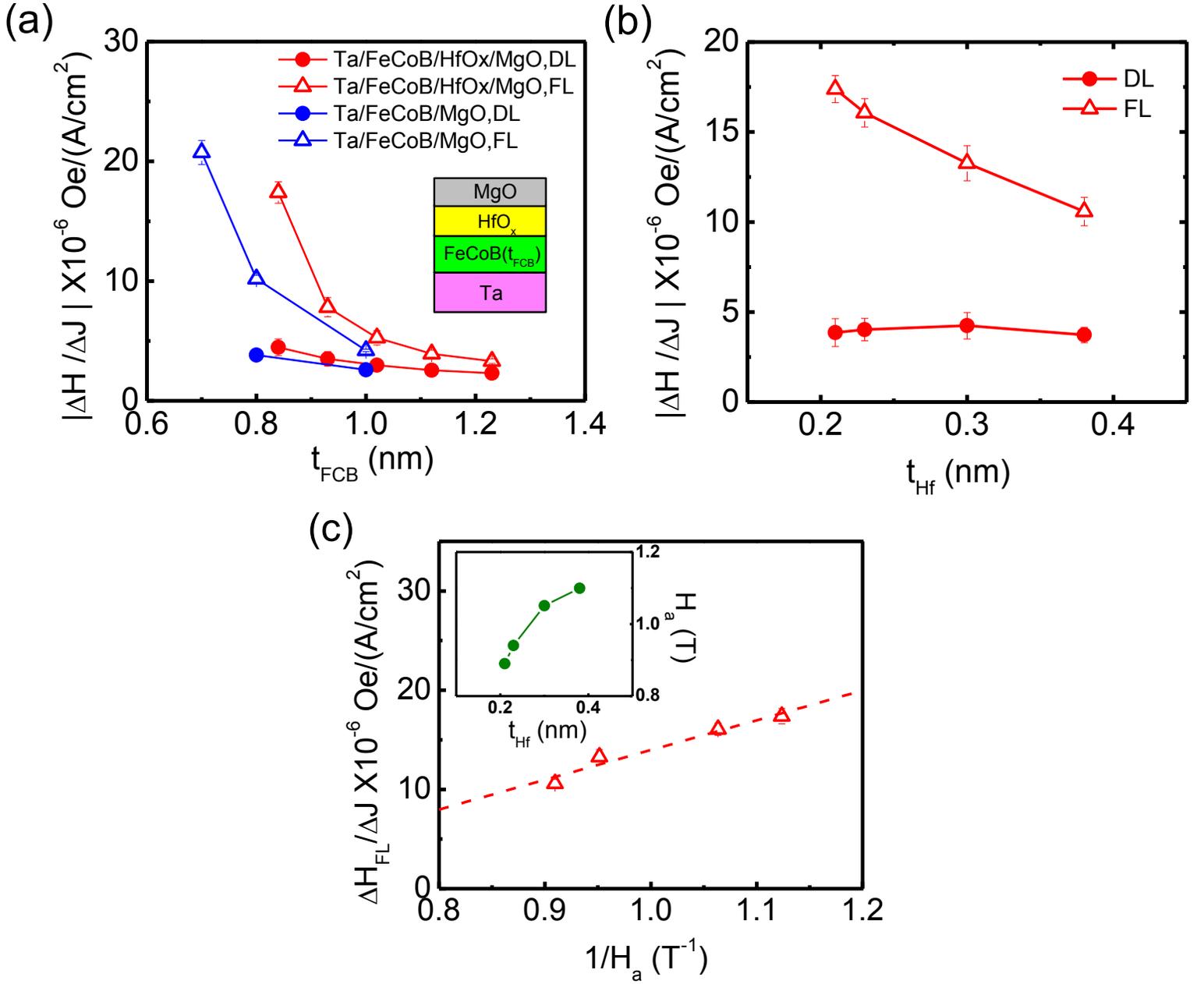

Fig. 2.(a) The current-induced effective fields $\Delta H_{DL}/\Delta J_e$ and $\Delta H_{FL}/\Delta J_e$ as a function of $t_{FeCoB}$ for a Ta/FeCoB/MgO and a Ta/FeCoB/HfO$_x$(2)/MgO sample. (b) $\Delta H_{DL}/\Delta J_e$ and $\Delta H_{FL}/\Delta J_e$ as a function of the Hf passivation layer thickness. (c) $\Delta H_{FL}/\Delta J_e$ as a function of the inverse of the anisotropy field $H_a$. Inset: $H_a$ for different Hf passivation layer thicknesses.

Fig 3

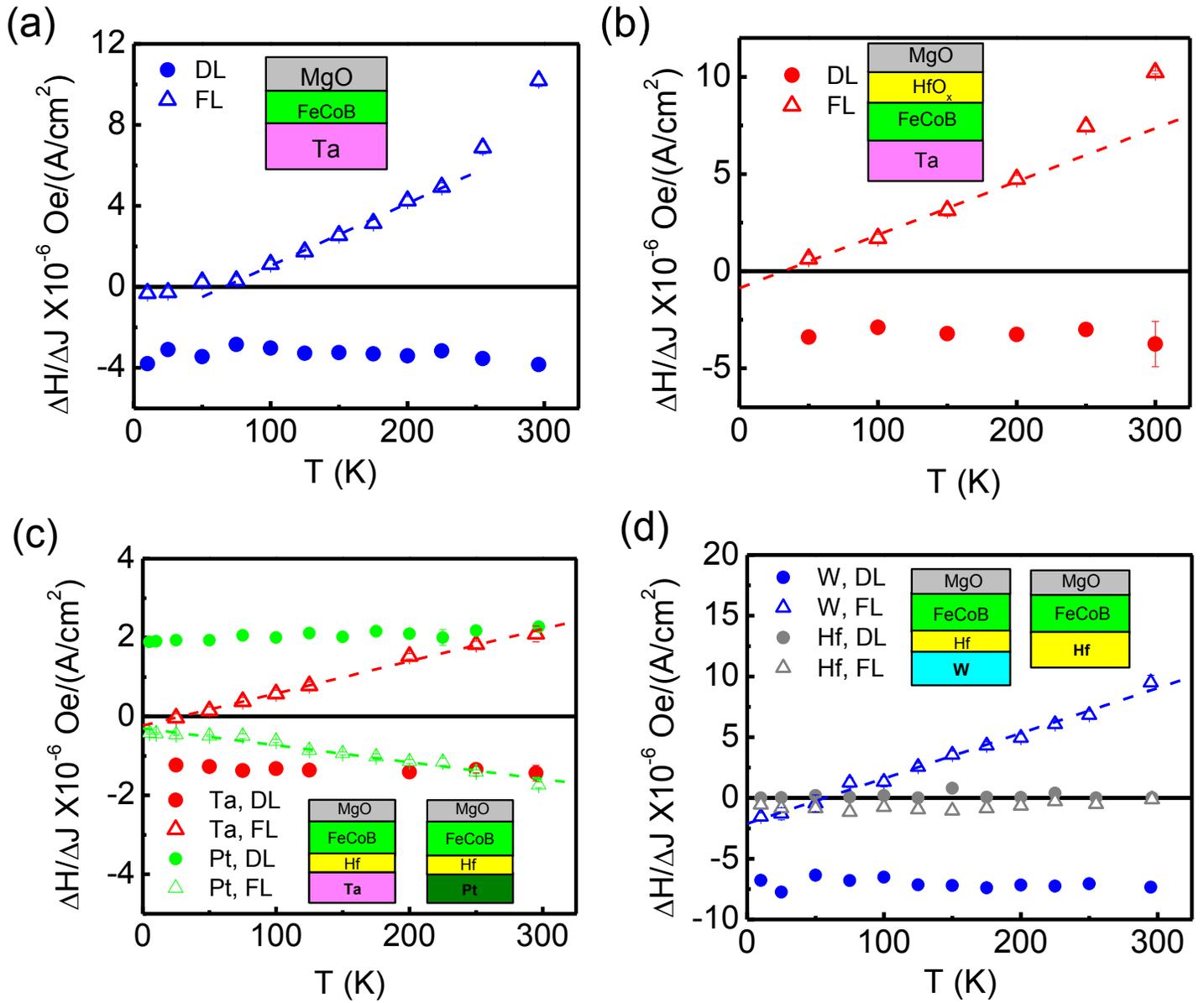

Fig. 3. Spin-orbit torque effective fields as a function of temperature for different samples: (a) Ta(4)/FeCoB(0.8)/MgO; (b) Ta(6)/FeCoB(0.8)/HfO$_x$(0.2)/MgO; (c) Ta(4)/Hf(1)/FeCoB(1)/MgO and Pt(4)/Hf(0.5)/FeCoB(1)/MgO; and (d) W(4)/Hf(1)/FeCoB(1)/MgO and Hf(4)/FeCoB(1)/MgO. The dashed lines are fits to the linear portion of the $\Delta H_{FL}(T)/\Delta J_e$ variation.

# Supplementary Material

## On the origin and location of the field-like spin-orbit torques in heavy metal/ferromagnet/oxide thin film heterostructures

Yongxi Ou[1], Chi-Feng Pai[1,*], Shengjie Shi[1], D. C. Ralph[1,2], and R. A. Buhrman[1,+]

[1]Cornell University, Ithaca, New York 14853, USA

[2]Kavli Institute at Cornell, Ithaca, New York 14853, USA

(*Present address: Department of Materials Science and Engineering, National Taiwan University, Taipei 10617, Taiwan)

[+]rab8@cornell.edu

## Contents

S1. Sample fabrication

S2. Analysis of spin diffusion in in-plane-magnetized (IPM) W/Hf/FeCoB/MgO samples

S3. Temperature dependent ST-FMR measurements on IPM W/Hf/FeCoB/MgO samples

S4. Magnetic characterization of perpendicularly magnetized (PM) Ta/FeCoB/Hf/MgO samples

S5. Transverse spin Hall magnetoresistance (TSMR) correction to the Harmonic response (HR) measurements

S6. HR measurement results from PM Ta/FeCoB/MgO samples having different annealing temperatures

S7. Temperature dependent HR measurement on PM Ta/FeCoB/MgO samples

## S1. Sample fabrication

We fabricated both in-plane-magnetized (IPM) and perpendicularly magnetized (PM) samples for this study. All the samples were fabricated via direct current (DC) sputtering (with RF magnetron sputtering for the MgO layer) in our deposition chamber with a base pressure $<8\times10^{-8}$ Torr. The DC sputtering conditions were 2mTorr Ar pressure, 30 watts power, with the following rates for the different elements: Ta: 0.014 nm/s, Pt: 0.012 nm/s, W: 0.014 nm/s, $Fe_{60}Co_{20}B_{20}$: 0.006 nm/s, Hf: 0.01 nm/s, MgO: 0.005 nm/s. The substrates were oxidized Si wafers $Si/SiO_2$.

The IPM samples reported on here were all W based: substrate/W(4)/Hf($t_{Hf}$)/Fe$_{60}$Co$_{20}$B$_{20}$($t_{FeCoB}$)/MgO(2)/Ta(1), where, for a given Hf thickness (0.25, 0.5, 1, 2nm), the FeCoB thickness was varied from 2 to 7nm. The numbers in parentheses represent the nominal thickness in nm. These samples were all annealed at 300 C for 1 hour under vacuum before measurement. For the ST-FMR measurements the IPM samples were patterned by photolithography and argon ion milling into microstrips and larger contact pads with the microstrip dimension being $20\ \mu m \times 10\ \mu m$.

The PM samples include samples with and without a Hf passivation layer inserted between FeCoB and MgO. For the samples with the Hf passivation layer, they are substrate/Ta(6)/FeCoB($t_{FeCoB}$)/Hf($t_{Hf}$)/MgO(2)/Ta(1), with $0.8\text{ nm} < t_{FeCoB} < 1.3\text{ nm}$. These samples all possessed strong perpendicular magnetic anisotropy (PMA) without any post-fabrication annealing process, beyond standard lithographic processing temperatures (=115 C). The Hf passivation layer had thicknesses ranging up to 0.4 nm, before the deposition of the MgO overcoat which results in the oxidation of the Hf, as determined by the strong anisotropy and by the order of magnitude increase in the resistance of MgO tunnel junction that had the Hf passivation layer, over that of junctions without the Hf.

The PM samples without the Hf passivation layer have different heavy metal base layers: substrate/Ta(4)/FeCoB($t_{FeCoB}$)/MgO(2)/Ta(1), SiO$_2$/Ta,Pt or substrate/W(4)/Hf($t_{Hf}$)/FeCoB(1)/MgO(2)/Ta(1) and substrate/Hf(4)/FeCoB(1)/MgO(2)/Ta(1). These samples were all annealed at 300 C for 1 hour under vacuum to enhance the PMA. All the PM samples were patterned by photolithography and argon ion milling into micrometer size Hall bars and larger contact pads with the dimension of the Hall bar being 60 $\mu$m×5 $\mu$m.

## S2. Analysis of spin diffusion in in-plane-magnetized (IPM) W/Hf/FeCoB/MgO samples

We use a trilayer spin diffusion model to describe the spin current and spin accumulation at the Hf/FeCoB interface when the spin current originates from the spin Hall effect (SHE) in the W underlayer. This schematic illustrates the model:

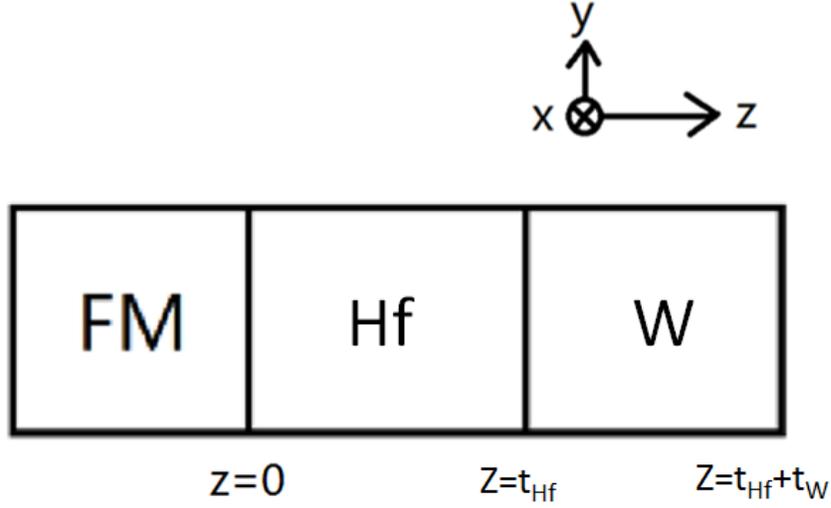

where we have assumed that the electrical current is flowing in the x direction.

The spin accumulation $\mu^s(z)$ in Hf and W layers must satisfy the diffusion equations:

$$\frac{\partial^2 \mu_{Hf}^s(z)}{\partial z^2} = \frac{\mu_{Hf}^s(z)}{\lambda_{s,Hf}^2} \quad \text{(S2.1)}$$

$$\frac{\partial^2 \mu_{W}^s(z)}{\partial z^2} = \frac{\mu_{W}^s(z)}{\lambda_{s,W}^2} \quad \text{(S2.2)}$$

where $\lambda_{s,Hf}$ ($\lambda_{s,W}$) is the spin diffusion length of the Hf(W) layer. Similarly, the spin current $j_s(z)$ in the two non-magnetic layers can be expressed as:

$$j_{s,\text{Hf}}(z) = -\frac{\sigma_{\text{Hf}}}{2e}\frac{\partial \mu^s_{\text{Hf}}}{\partial z} \qquad (S2.3)$$

$$j_{s,\text{W}}(z) = -\frac{\sigma_{\text{W}}}{2e}\frac{\partial \mu^s_{\text{W}}}{\partial z} - j^{\text{SHE}}_{\text{W}} \qquad (S2.4)$$

Here $j^{\text{SHE}}_{\text{W}}$ is the spin current generated by the SHE in the W and we have assumed that the spin current originating from the amorphous Hf layer is negligible, which is confirmed by our harmonic response measurements (see Fig. 3(d) in the main text).

We assume the following boundary conditions:

$$j_{s,\text{W}}(t_{\text{Hf}} + t_{\text{W}}) = 0 \quad \text{(spin current vanishes at the metal/substrate interface)} \qquad (S2.5)$$

$$j_{s,\text{Hf}}(t_{\text{Hf}}) = j_{s,\text{W}}(t_{\text{Hf}}) \quad \text{(spin current continuity at the W/Hf interface)} \qquad (S2.6)$$

$$\mu^s_{\text{Hf}}(t_{\text{Hf}}) = \mu^s_{\text{W}}(t_{\text{Hf}}) \quad \text{(spin accumulation continuity at the W/Hf interface)} \qquad (S2.7)$$

$$j_{s,\text{Hf}}(0) = -\frac{\text{Re}\,G^{\uparrow\downarrow}}{e}\mu^s_{\text{Hf}}(0) \quad (\text{Re}\,G^{\uparrow\downarrow}: \text{spin mixing conductance at the Hf/FeCoB interface}) \qquad (S2.8)$$

By combining Equations (S2.1) to (S2.8), we can express the spin accumulation and spin current at the Hf/FeCoB interface as:

$$\mu^s_{\text{Hf}}(0) = \frac{K_{\text{Hf}}\,j^{\text{SHE}}_{\text{W}}}{e^{t_{\text{Hf}}/\lambda_{\text{Hf}}} + P_{(\text{Hf},\text{W})}e^{-t_{\text{Hf}}/\lambda_{\text{Hf}}}} \qquad (S2.9)$$

$$j_{s,\text{Hf}}(0) = -\frac{\text{Re}\,G^{\uparrow\downarrow}}{e}\mu^s_{\text{Hf}}(0) \propto \frac{K_{\text{Hf}}\,j^{\text{SHE}}_{\text{W}}}{e^{t_{\text{Hf}}/\lambda_{\text{Hf}}} + P_{(\text{Hf},\text{W})}e^{-t_{\text{Hf}}/\lambda_{\text{Hf}}}} \qquad (S2.10)$$

Here we have defined:

$$K_{Hf} = \frac{4eG_{Hf}}{(2G^{\uparrow\downarrow} + G_{Hf})(G_{Hf} + G_W)} \quad \text{and} \quad P_{(Hf,W)} = \frac{2\operatorname{Re}G^{\uparrow\downarrow} - G_{Hf}}{2\operatorname{Re}G^{\uparrow\downarrow} + G_{Hf}} \cdot \frac{G_{Hf} - G_W}{G_{Hf} + G_W}$$

where the spin conductance of Hf(W) is defined as: $G_{Hf(W)} = \sigma_{Hf(W)} / \lambda_{Hf(W)}$. From equation (S2.9) and (S2.10), one can see that both the spin current and spin accumulation have the same $t_{Hf}$ dependence: $1/(e^{t_{Hf}/\lambda_{Hf}} + P_{(Hf,W)} e^{-t_{Hf}/\lambda_{Hf}})$. We have used this function to fit the data in Fig. 1(b) in the main text, where we used $\rho_W \approx 200 \mu\Omega \cdot cm$, $\rho_{Hf} \approx 120 \mu\Omega \cdot cm$ (from resistance measurements), $\lambda_W = 2.5$ nm, $\lambda_{Hf} = 1$ nm, $\operatorname{Re}G^{\uparrow\downarrow} = 1 \times 10^{15}\ \Omega^{-1} m^{-2}$. These values yields $P_{(Hf,W)} \approx 0.25$, which indicates that the dominant term in $1/(e^{t_{Hf}/\lambda_{Hf}} + P_{(Hf,W)} e^{-t_{Hf}/\lambda_{Hf}})$ is $1/e^{t_{Hf}/\lambda_{Hf}}$. The fit in Fig. 1(b) gives a self-consistent value $\lambda_{Hf} = 0.9 \pm 0.2$ nm. Notice that in order to fit the FL term, we include a constant offset in the expression $\xi_{FL,0}$, namely:

$$|\xi_{FL}(t_{Hf})| = \frac{|\xi_{FL}(0)|}{e^{t_{Hf}/\lambda_{Hf}} + P_{(Hf,W)} e^{-t_{Hf}/\lambda_{Hf}}} + |\xi_{FL,0}| \qquad (S2.11)$$

The fit gives $|\xi_{FL,0}| = 0.010 \pm 0.004$ (with negative sign), which we attribute to a small Rashba-Edelstein effect at the Hf/FeCoB interface that yields a residual field-like torque in the absence of any spin accumulation from SHE.

## S3. Temperature dependent ST-FMR measurements on IPM W/Hf/FeCoB/MgO samples

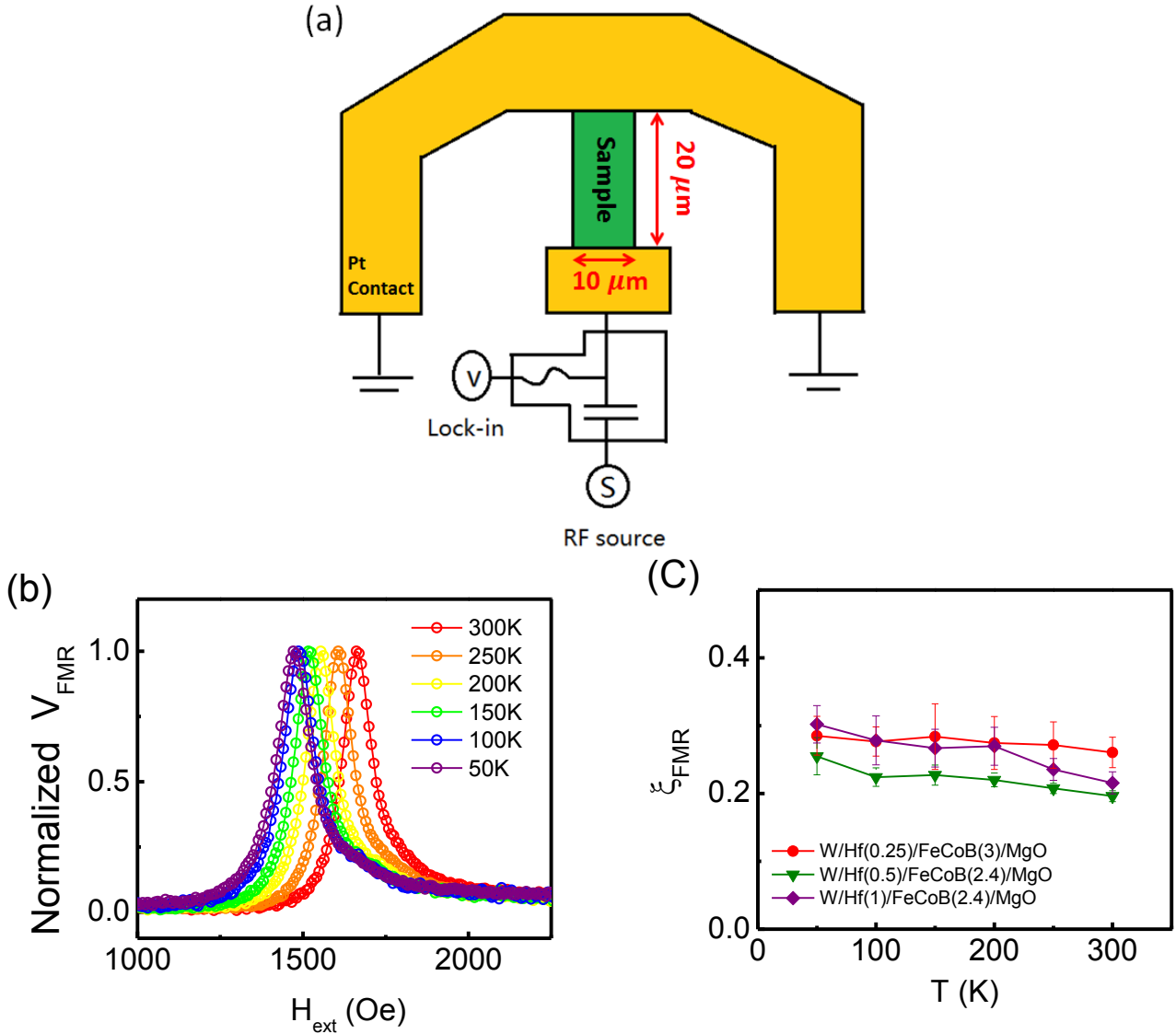

Fig. S3.(a) Schematic of the ST-FMR measurement. (b) ST-FMR voltage signal from variable temperature measurements on a sample W(4)/Hf(1)/FeCoB(2.4)/MgO at frequency 11 GHz. (c) ST-FMR spin torque efficiency $\xi_{FMR}$ as a function of temperature for samples W(4)/Hf(0.25, 0.5, 1)/FeCoB(2.4, 3)/MgO.

In order to study the temperature dependence of the FL torque in the W/Hf/FeCoB/MgO IPM samples, we performed ST-FMR on these samples in a flow cryostat with liquid helium cooling. The schematic of the measurement is shown in Fig. S3(a). Figure S3(b) shows results for a IPM sample W(4)/Hf(1)/FeCoB(2.4)/MgO at a fixed frequency 11 GHz. The line shape of the voltage signals maintains the same shape, as can be seen from Fig. S3(b). Figure S3(c) shows the calculated spin torque efficiency $\xi_{FMR}$, which is a ratio of the symmetric component to the antisymmetric component to the voltage [1,2], as a function of temperature for samples W(4)/Hf(0.25, 0.5, 1)/FeCoB(2.4, 3)/MgO. The variation of $\xi_{FMR}$ is small. Previous studies [3,4] and our results (e.g. Fig. 3(d) in the main text) both indicate that the DL torque is not sensitive to temperature. As we discussed in the main text, the FL torque in IPM samples is attributable to the metal/FM interface. The small variation of $\xi_{FMR}$ indicates that the FL torque at the metal/FM interface also only has a weak temperature dependence, which is strikingly different from the FL torque that is exerted at the FM/Oxide interface as observed in the PM samples.

## S4. Magnetic characterization of perpendicularly magnetized (PM) Ta/FeCoB/Hf/MgO samples

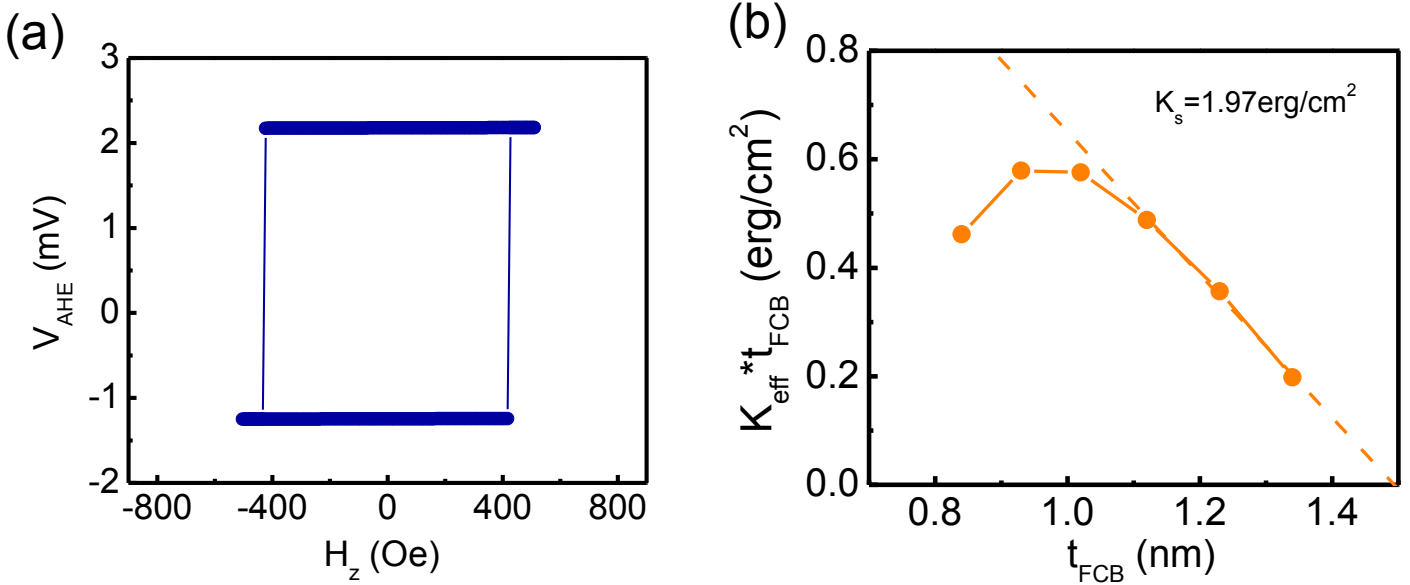

Fig. S4.(a) $V_{AHE}$ versus $H_z$ for the sample Ta(6)/FeCoB(1)/Hf(0.2)/MgO(2)/Ta(1). (b) $K_{eff}*t_{FeCoB}$ as a function of $t_{FeCoB}$ for samples Ta(6)/FeCoB($t_{FeCoB}$)/Hf(0.2)/MgO(2)/Ta(1). The dashed line is the linear fit to the thicker samples, from which the intercept gives $K_i \simeq 2\,\text{erg/cm}^2$ according to equation (S4.1).

As described in Section S1, the PM Ta/FeCoB/Hf/MgO samples have a Hf passivation layer that becomes oxidized during the subsequent sputtering of the MgO layer. These samples possess PMA without further post-fabrication annealing with a typical anisotropy field $H_a \sim 1\,\text{T}$. Figure

S4(a) shows a typical loop of the anomalous Hall voltage $V_{AHE}$ versus perpendicular field $H_z$ for a sample Ta(6)/FeCoB(1)/Hf(0.2)/MgO(2)/Ta(1) indicating that the sample has a coercive field of ~400 Oe. In order to characterize the interfacial PMA, we perform harmonic response measurements on a PM sample Ta(6)/FeCoB($t_{FeCoB}$)/Hf(0.2)/MgO(2)/Ta(1) to attain the anisotropy field and estimate the effective anisotropy energy $K_{eff}$ [5]:

$$K_{eff} = K_b + \frac{K_i}{t_{FM}} \quad (S4.1)$$

where $K_b(K_i)$ is the bulk(interfacial) anisotropy energy term. In Figure S4(b) we plot $K_{eff} * t_{FeCoB}$ versus $t_{FeCoB}$, from the zero thickness intercept to the fit to the linear region of this plot we estimate the interfacial anisotropy energy $K_i \simeq 2\,\text{erg/cm}^2$.

## S5. Transverse magnetoresistance (TMR) correction to the harmonic response (HR) measurements

Previous work has shown that when analyzing the results of HR measurements, it is necessary to correct for the transverse (planar) Hall magnetoresistance contribution to the measurement signal [6]. If the transverse resistance is $R_{THM}$, and the anomalous Hall resistance is $R_{AHE}$, then the ratio $\delta = R_{THM}/R_{AHE}$ will enter into the correction as [6]:

$$\Delta H_{DL} = \frac{\Delta H_L + 2\delta \cdot \Delta H_T}{1 - 4\delta^2}$$
$$\Delta H_{FL} = \frac{\Delta H_T + 2\delta \cdot \Delta H_L}{1 - 4\delta^2} \quad (S5.1)$$

where $\Delta H_L$ ($\Delta H_T$) is the longitudinal (transverse) measured field when sweeping the external field along (transverse to) the current direction.

The transverse Hall magnetoresistance (THM) can include a contribution from both the anisotropic magnetoresistance of the FM layer, as well as from the spin Hall magnetoresistance that arises from the combination of spin accumulation and the inverse spin Hall effect in the HM [7]. Both contributions vary with the product of the two in-plane magnetic moment components ($m_x m_y$). All of our samples discussed in the main text have $|\delta| < 0.16$, so that the correction Eq. (S5.1) is small for our samples. Since the temperature dependence of $\delta$ has been reported as being small for systems similar to those we study here [3,4], we have used the room temperature values of $\delta$ for our variable temperature HR results.

## S6. HR measurement results from PM Ta/FeCoB/MgO samples having different annealing temperatures

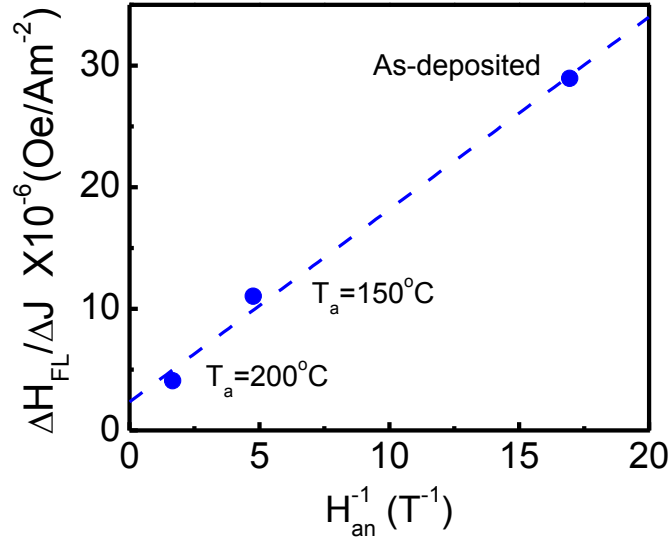

Fig. S5. $\Delta H_{FL} / \Delta J_e$ as a function of $H_{an}^{-1}$ for a sample Ta(4)/FeCoB(1)/MgO after annealing at different temperatures.

In the main text, we discuss how we could alter the interfacial PMA by inserting thin Hf passivation layers of different thickness between the FeCoB and MgO. As an independent experiment, we also varied the PMA of a different set of samples by annealing a PM Ta(4)/FeCoB(1)/MgO sample at different temperatures for 1 hour in each case. The as-deposited sample shows a small anisotropy field $H_{an} = 590\,\text{Oe}$, while annealing at 150 C° and 200 C° resulted in $H_{an} = 2100\,\text{Oe}$ and $H_{an} = 6000\,\text{Oe}$ respectively. We measured the FL effective field per current density $\Delta H_{FL} / \Delta J_e$ via the harmonic response technique, and plot the results as a function of $H_{an}^{-1}$ as shown in Fig. S5. The dashed line is the linear fit to the data. Again, the results here are consistent with the observation obtained from the Hf passivation layer samples (Fig. 2(c) in the main text), indicating an inverse correlation between the strength of the FL torque at the FeCoB/Oxide interface and that of the interfacial magnetic anisotropy.

## S7. Temperature dependent HR measurement on PM Ta/FeCoB/MgO samples

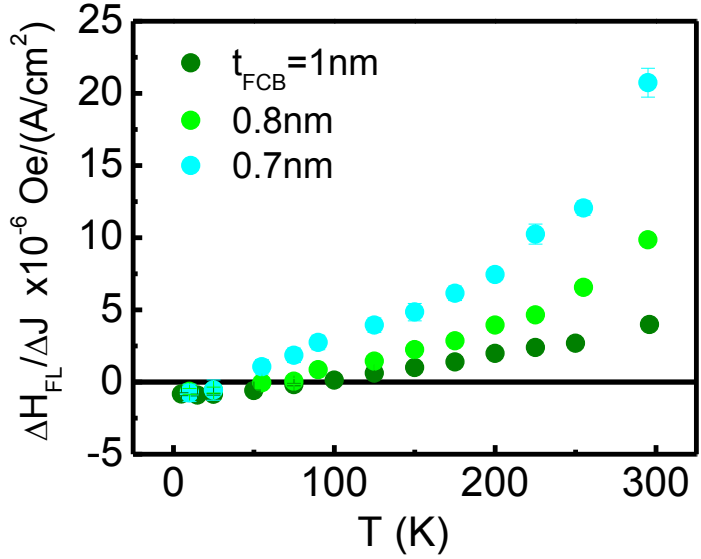

Fig. S6. $\Delta H_{FL} / \Delta J_e$ as a function of temperature T for samples Ta(4)/FeCoB(0.7, 0.8, 1)/MgO.

Fig. S6 shows the temperature dependence of the FL effective field $\Delta H_{FL} / \Delta J_e$ as obtained for a set of PM samples Ta(4)/FeCoB($t_{FeCoB}$)/MgO where $0.7\text{nm} \leq t_{FeCoB} \leq 1\text{nm}$. The estimated Oersted field contribution has been subtracted from the data and the positive sign of $\Delta H_{FL}$ indicates that it is oriented parallel to the Oersted field. As shown in Fig. S6, $\Delta H_{FL}$ at room temperature (RT) varies by nearly an order of magnitude while the thickness of the FeCoB layer varies by only a small amount (30%). We conclude that this is the principally the result of a larger portion of the SHE-generated spin current reaching the FeCoB/MgO interface when the FeCoB layer is thinner. In these samples the perpendicular anisotropy field $H_a$ varied only by 18% with this thickness variation. We note that relatively large values of $\Delta H_{FL}$ have also been observed in similar of Ta/FeCoB/MgO heterostructures for ultrathin FeCoB samples at RT in previous studies [3,4,6]. Even though $\Delta H_{FL}$ can vary by nearly an order of magnitude at RT for different FeCoB thicknesses, the results for all three samples converge a very small (negative) value as the sample temperature approaches 5 K, consistent with the temperature dependent behavior found in previous studies of PM Ta/FM/MgO samples [3,4], and indicative, we suggest, of a freezing out of the spin relaxation process at the MgO interface states at low temperature, leaving only the field-like torque that is generated at the Ta/FM interface.